# Microcanonical Molecular Dynamic Simulations of Au Nanoclusters

Karina L. D. Barturen H.[(*)],  F. A. R. Navarro - Universidad Nacional de Educación,

Justo Rojas T. - Universidad Nacional Mayor de San Marcos

Lima – Peru

## Abstract

In this paper, we study nanoparticles with constituent atoms ranging from dozens to hundreds of them. These types of particles display structural and magnetic properties that strongly depend on the number of constituents *N*. The metal clusters are important due their interesting properties when compared to bulk materials; hence they have potential technological applications. Specifically, we study the Au nanoclusters through classical molecular dynamics simulations; we analyze the total and potential energy as a function of time. Likewise, we study the geometrical structures of Au Nanocluster corresponding to the lowest energy states at 0 K. We consider the method of microcanonical ensemble, and we carry out computer simulations by operating the XMD software package and the atomistic configuration viewer AtomEye.

## 1. Introduction

The metal clusters are intermediate states of matter among single atoms and bulk. They have tiny dimensions and a high surface/volume ratio. Those nanoparticles have an average sizes between 20 and 100 nm, spherical shapes and size dispersion extremely low. We should stress that the structural and electronic properties of the clusters depend strongly on their size [1]. The research about these clusters is huge because the practical applications include their usage as catalysts and as the basement for the incoming nanoelectronics [2, 3].

On 2002, Wilson Ho *et al*. discovered the molecular phase [4, 5] from which a group of single Au atoms become a solid structure, that is, a one-dimensional gold chain. They done experiments with the scanning tunneling microscope, and from single atoms they attained to building gold chains on NiAl(110). Their studies suggest a minimum size limit for the construction of electronically conductive molecules. On 2005, A. Calzolari and M. B. Nardelli utilize first-principle calculations for researching the stable phases of Au chains on the (110) surface of NiAl [6].

In scientific literature there are researches utilizing molecular dynamics simulations for studying nanoparticles, namely, Chunjie Yang *et al*. operated it for investigating positively charged silver nanoparticles [7]. Also, Dong Hwa Seo *et al*. took into account molecular

[(*)] corresponding author, **karinabart@gmail.com**

dynamics simulation for probing the diffusion of Au and Pt nanoclusters on carbon nanotubes [8].

## 2. Theoretical Frame

The molecular dynamics is a powerful technique that yields information about the simulated system. It lies, essentially, in integrating the Newton's equations of Motion through numerical methods [9, 10]. Below, briefly, we will display a standard algorithm for a molecular dynamics simulation. The information generated by a run of molecular dynamics is the position and the velocity of each particle at each instant of time. We can, by utilizing statistical mechanics, pass from this microscopic information to macroscopic magnitudes. These latter allow us to contrast with actual outcomings. The conventional ensemble for molecular dynamics is the microcanonical one. The molecular dynamics allows obtaining successive configurations in the time. So, we can calculate dynamic properties such as spatial and temporal correlations, diffusion coefficient, etc. Then, we display the five basic ingredients that we should take into account into an algorithm for a molecular dynamics simulation:

> **1.** Composition of the system:
> - Define the number, kind and mass of the atoms.
> - Define the atomic interactions.
>
> **2.** Initial conditions at t = 0
> - Give the positions and velocities of particles.
>
> **3.** Definition of the time step $\Delta t$.
>
> **4.** Simulation process
> - Calculation of the forces at $t = 0$.
> - Calculation of the positions at $t + \Delta t$.
> - Calculation of the forces at $t + \Delta t$.
> - Calculation of the velocities at $t + \Delta t$.
> - Increasing of the time at $\Delta t$.
>
> **5.** Analyze results.

,

## 3. Computer simulations

By utilizing classical molecular dynamics, we achieved various geometric structures of Au nanocluster. Theoretically, we know that the clusters take geometric structures in accordance to the local minima on the potential energy surface. So, it is expected that the clusters take a number of stable configurations as the number of local minima. To finding out the global minimum, the program, for each cluster, run starting from different initial configurations. We obtained an energy spectrum to both zero and equilibrium temperature. We notice the more the total number of particles increases the more the number of local minima increase [11, 12].

The Figures 1 and 2 display some of the most stable structures of nanoclusters Au $_n$, for n = 5-16 atoms. We viewed it through the program AtomEye [13]. We notice that the structures, corresponding to the minimum energy obtained in the $Au_{10}$ cluster simulation, have a well defined regular geometry. Specially, for $Au_{13}$ the most stable structure is a regular icosahedron.

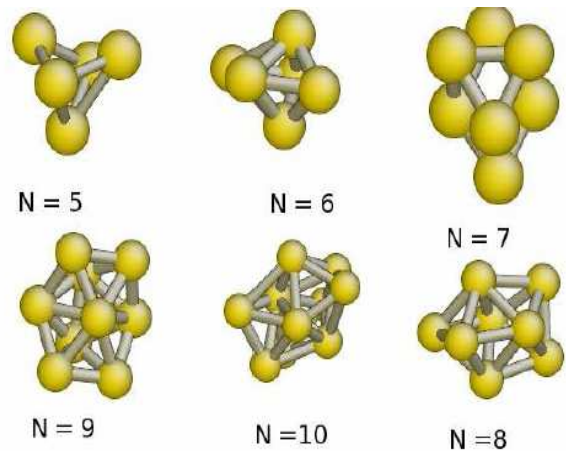

**Fig.1.** For T=0, the more stables geometrical structures of clusters Au $_n$ (n=5-10).

,

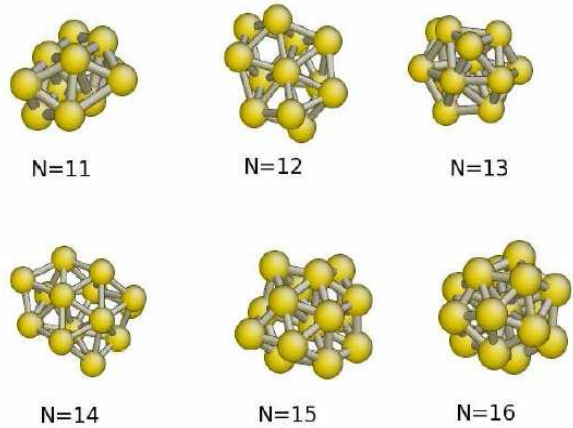

**Fig.2.** For T=0, the more stables geometrical structures of clusters Au $_n$ (n=11-16).

In the Table 1 are calculated values of the potential energies of the relatively more stable nanocluster in the Au $_n$ system. The greater the atom number in the cluster the more the potential energy diminishes.

**Table 1.** Bond energy of Au $_n$ clusters, with n=4-10.

| $n$ | $-E_p(eV)$ | n | $-E_p(eV)$ |
|---|---|---|---|
| 4 | 2.5859 | 11 | 3.07984 |
| 5 | 2.74362 | 12 | 3.10596 |
| 6 | 2.82917 | 13 | 3.13135 |
| 7 | 2.92674 | 14 | 3.15555 |
| 8 | 3.00326 | 15 | 3.16003 |
| 9 | 3.04593 | 16 | 3.22696 |
| 10 | 3.03031 | 17 | 3.17470 |

In the Figure 3, we see the $Au_{10}$ cluster displays an increasing in potential energy; it means this cluster is less stable when compared to its neighbors. Even, the $Au_{16}$ cluster is more stable when contrasted to the $Au_{15}$ and $Au_{17}$ clusters.

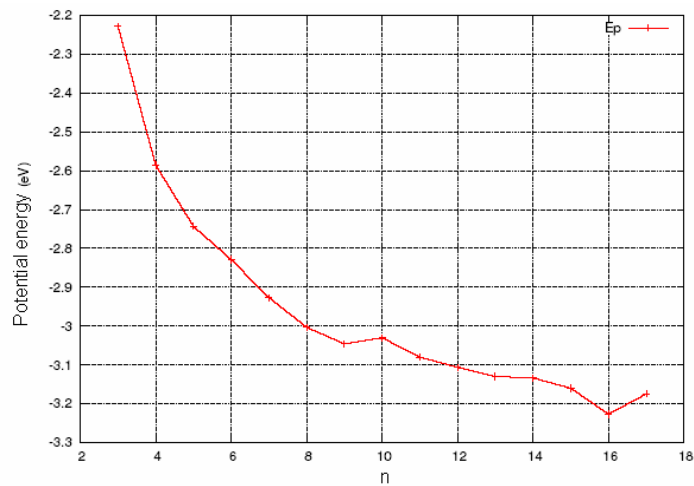

Fig.3. Potential energy as a function of atom number into cluster.

In the Figure 4, we display the variation of total energy and potential energy for the Au $_n$ nanocluster, with n=16, as function of the time steps.

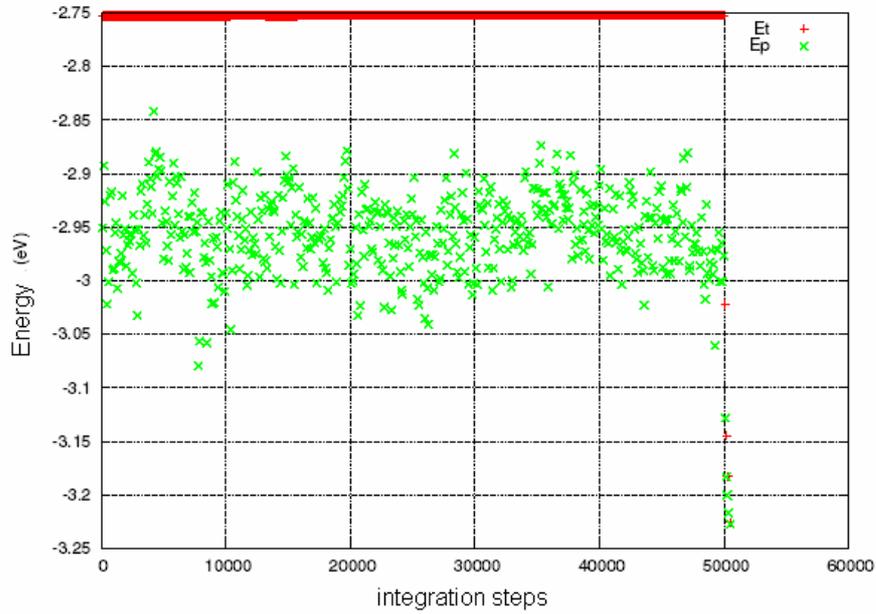

**Fig.4.** Variations of total Energy (+) and potential energy (*) for the Au $_n$ nanocluster, with n=16, as function of the time steps.

**3.1 Bond Length in the Au Clusters**

We have also calculated the external average interatomic spacing in the cluster. They are displayed in Table 2. In the majority of cases analyzed it was observed the average lengths of the internal bond between Au-Au atoms are larger than average lengths of external bond, this is correct for n greater than 13. Besides, the greater the particles number the greater that distance. In the case of $Au_{10}$ nanoparticle the average length of external bond is 2.661 Å, that is, higher than average lengths of the external bond of Au $_n$ atoms, n= 4-14. However, the average length of external bonds is less for Au $_n$ atoms with n=15-16. In all cases the average lengths are smaller than the corresponding Au bulk; the minimum interatomic distance in FCC Au bulk is 2.88 Å.

**Table 2.** Average length (in Å): for internal bonds (int) and external bonds (ext) among Au-Au atoms at 0 K.

| $n$ | $(Au-Au)_{int}()$ | $(Au-Au)_{ext}()$ |
|---|---|---|
| 4 | - | - |
| 5 | - | 2.503 |
| 6 | - | 2.547 |
| 7 | - | 2.568 |
| 8 | - | 2.597 |
| 9 | - | 2.618 |
| 10 | - | 2.661 |
| 11 | - | 2.636 |
| 12 | - | 2.635 |
| 13 | 2.691 | 2.727 |
| 14 | 2.713 | 2.642 |
| 15 | 2.800 | 2.679 |
| 16 | 2.838 | 2.699 |

## 4. Conclusions

We applied classical molecular dynamics by using the XMD software for studying Au nanoclusters. We also utilized the known atomistic configuration viewer AtomEye for knowing their geometrical structures corresponding to the lowest energy states at 0 K. We have found that structural and magnetic properties strongly depend on *N*, the number of constituent atoms in the metal cluster. In other words, we have attained as the relatively more stable structure as the respective energies of Au $_n$ nanoclusters, for n= 4-16.

,